\documentstyle  [12pt]{article}
\newcommand{\eqn}{\begin{equation}}
\newcommand{\eqnend}{\end{equation}}

\begin{document}
\baselineskip 0.8truecm

\title{Bend conductance of crossed wires in the presence of Andreev
scattering}

\author{ Simon J Robinson \\
\\
 School of Physics and Materials \\
 University of Lancaster, Lancaster, LA1 4YB, UK }

\date{}

\maketitle

\begin{abstract}

We study the 4-probe  bend conductance $G_{14,32}$
 of a mesoscopic crossed wire structure
in the ballistic regime in the absence of  a magnetic field,
which for normal devices is usually negative.
We
predict that for sufficiently large devices and for small applied voltages
 $G_{14,32}$ becomes positive in the presence of Andreev scattering,
but at large voltages  $G_{14,32}$ remains negative.

\end{abstract}


\pagebreak

Although the theory of transport in mesoscopic systems connected in
multiprobe arrangements~\cite{buttiker:multi86,buttiker:multi88}
is now well understood and has been
successfully applied to a number of situations, the equivalent analysis
for a superconductor has only recently been developed~\cite{here:multi93}.
Using
the results of ref.~\cite{here:multi93} it has been
shown~\cite{here:negative94} that the sign of the
longitudinal 4-probe conductance of a disordered wire, which would be
expected to be positive for normal materials, can be reversed if
the wire becomes superconducting.
In this letter we study another geometry  commonly used for multiprobe
measurements: The ballistic crossed wire arrangement of fig. 1, for
which the bend conductance is defined as $G_{14,23}$ where
$G_{ij,kl}=I_{i}/(V_k-V_l)$ subject to the constraint $I_{j}=-I_{i}$,
$I_k=I_l=0$, and $I_m$ is the current flowing into the sample
from reservoir $m$. We will assume the normal system is smaller
than the elastic and inelastic
scattering lengths and study the effect of introducing
superconducting order parameter $\Delta$.
We will show that  in the absence of a magnetic field, the sign of
 $G_{14,23}$ depends on whether the dominant transmission process is
across the device or round the
 corner (bend transmission). In the presence of superconductivity
the bend transmission can dominate for quasi-particle energies below the
gap, whereas for high quasi-particle energies and for normal materials
this is not the case.  We deduce that for small applied voltages
 the bend conductance
can change sign if the device becomes superconducting and the width $W$
of the junction is sufficiently large. If this occurs, the sign may reverse
again as the voltage is increased.
The analysis is carried out at zero temperature.

The case of a normal material in a zero or weak magnetic field was
studied by Avishai and Band~\cite{avishai:res89}. The presence of a strong
magnetic field corresponds to the quantum hall regime, and the
bend conductance is known to be large~\cite{baranger:bend91}.
We study here the case of
no
applied field, and following ref~\cite{avishai:res89}
will assume that the system
is symmetric so that it is possible to characterise it completely in
terms of only three normal and three Andreev
 scattering probabilities. We will write
$T_{ij}^X=\sum_{mn} T_{im,jn}^X$ where $T_{im,jn}^X$
 is the transmssion probability for a particle incident on
the device from channel $n$ of probe $j$ to be transmitted to
channel $m$ of probe $i$, and $X=O$ for normal transmission, $A$ for
Andreev transmission. For the case $i=j$ we will interchangeably
substitute $T_{ii}^X=R_i^X$. For normal devices, $T_{ij}^A=0$.
Then we assume the probabilities of a particle being reflected,
transmitted round a corner, or transmitted straight across the junction are
independent of the probe along which it arrives, so that we
need only consider the probabilities
$R_1^X$,  $T_{31}^X$  and
$T_{21}^X$. For a normal
system  $R_1^O + T_{21}^O + 2T_{31}^O =N$
where $N$  is the number of channels in each probe and
it was found in~\cite{avishai:res89}
that~\cite{note}
   \eqn G_{14,32} = \frac{8e^2}{\hbar}
      \frac {T_{31} (T_{21}^O+T_{31}^O) }  {T_{21}^O-T_{31}^O}
       \label{eq:norg}
   \eqnend
$G_{14,32}$ was found in~\cite{avishai:res89} to be negative, in agreement with
experiment~\cite{takagaki:multi88}.
This was interpreted as showing that
$T_{21}>T_{31}$, and is consistent with numerical evidence from other
sources~\cite{wang:pc92,takagaki:4probe94} that for normal ballistic systems
electrons
travelling into regions
containing junctions will tend to travel in  straight lines rather
than scattering round corners.  We  will now show that this result
can be reversed in the presence of superconductivity. We consider the
situation of fig. 1, in which the order parameter $\Delta$ is
constant and nonzero in the shaded area, but zero everywhere else.
We note that quasi-particles below the energy gap will now decay
in the superconducting region as $\exp -x/\lambda$, where $\lambda$
varies from infinity for quasi-particle energies at the edge of
the gap to $E_F/\Delta k_F$ for quasi-particles at the Fermi energy.
{}From figure 2 it is clear that particles being transmitted around the
bend must travel through a length $2L$ of superconductor, while
those travelling straight through must travel through at least a length
equal to the smaller of $2L+W$  and $4L$. Hence if the conductance
measurement is made at zero temperature at voltages below
the gap and lengths $L$ and $W$ are both  greater than
 $E_F/\Delta k_F$, we expect that the bend transmission coefficients will
become exponentially larger than the straight-through probabilities.

We now consider the opposite limit, for which the superconducting order
parameter is sufficiently weak that it is possible to study the
scattering properties of the device using perturbation theory. This
can occur if  the length of the superconducting region
is smaller than
$E_F/\Delta k_F$ or the energy of incident quasi-particles is
sufficiently far above the superconducting energy gap.  In this case, we
would expect that switching on the superconductivity
would not substantially alter the relative probabilities of normal
transmission
accross the device and round a corner, but will introduce small
Andreev transmission coefficients. Hence if the dominant transmission
process in the normal material is straight across the device, the same
will be true in presence of superconductivity.

Having established criteria likely to cause transmission across the device
or around the bend to dominate, we now examine the implications for
the bend conductance.
 It was
shown in~\cite{here:multi93} that
   \eqn G_{ij,kl} = (2e^2/h) d/(b_{ik} + b_{jl} - b_{il} - b_{jk})
   \label{eq:multi} \eqnend
where $d$ is the determinant of a transport matrix $A$, $b_{ij}$
is the determinant of the $ij$'th cofactor of $A$, and $A$ is
given by $a_{ij} = N \delta_{ij} + T_{ij}^A - T_{ij}^O$.
We put
$R=N+R_1^A-R_1^O = 2R_1^A + T_{21}^A + T_{21}^O + 2t_{21}^A + 2t_{21}^O$,
  $T=T_{13}^A-T_{13}^O$
and $t=T_{12}^A-t_{12}^O$, and hence write
   \eqn A = \left( \begin{array}{cccc}
     R & T & t & t \\
     T & R & t & t \\
     t & t & R & T \\
     t & t & T & R  \end{array} \right)
   \eqnend
Note that despite the
notation $R$ and $T$ are not direct reflection and transmission
probabilities.  We immediately deduce
   \eqn d = (R-T)^2 (R +T + 2t) (R + T - 2t) \eqnend
   \eqn b_{13}=b_{42}=2 RTt-R^2 t - T^2 t  \eqnend
   \eqn b_{43}=b_{12}=T^3 + 2Rt^2 - R^2T - 2Tt^2 \eqnend
And hence
   \begin{eqnarray} G_{12,34} & = &
         \frac{e^2}{2\hbar} \frac{(R-T)(R+T-2t)}{T-t} \nonumber \\
                & & \nonumber \\
                & = & \frac{e^2}{2\hbar}
                  \frac{(R_1^A+T_{21}^O+T_{31}^A+T_{31}^O)
                            (R_1^A+T_{21}^A+2T_{31}^O)}
                       {T_{21}^A+T_{31}^O -T_{21}^O - T_{31}^A}
   \label{eq:posg} \end{eqnarray}
Expression~(\ref{eq:posg}) is the key result of this
paper. It shows that although in the presence of
Andreev scattering, $G_{14,32}$ is a relatively complex function of the
various scattering probabilities, its sign depends in a simple way
on whether particles tend to be transmitted across the junction or round
the corner, and on whether normal or Andreev transmission dominates.
Numerical
simulations~\cite{here:local90,here:mythesis92} have shown that in the absence
of magnetic effects
and disorder-induced strong localization,
as a rule of thumb the normal scattering coefficient will be the larger one.
Hence
we conclude that in this limit $G_{14,32}$ is positive if
$T_{31}^O > T_{21}^O$, and negative otherwise.  As a result if a normal
device with $T_{21}^O > T_{31}^O$ becomes superconducting such that
$E/\Delta k_F < L,\; W$ then at zero temperataure and for small voltages
the bend conductance will reverse sign and become positive.
 At finite tempeatures or voltages,
expressions~(\ref{eq:multi}-\ref{eq:posg}) remain valid but the
coefficients $T_{ij}^X$ must now be interpreted as energy integrals over the
range of energies at which  quasi-particles are incident on the
device, as explained in
ref~\cite{here:multi93}. For applied voltages greater than the
energy gap $\Delta$, the incident quasi-particles at the highest energies
will have high transmission probabilities and tend to be transmitted
across the device. Since these probabilities will form the dominant
contribution to $T$ and $t$, the denominator of~(\ref{eq:posg}) will
become negative. Hence we predict that for sufficiently high voltages
$G_{14,32}$ will become negative.

Although the above analysis was based on the assumption of perfect symmetry,
the conclusions are unchanged if we allow small asymmetries to be present
so that the channels are not identical. This is not a trivial point,
since we need to exclude the possibility of  terms of order of magnitude
of the differences between the different reflection coefficients
appearing in the denominator of~(\ref{eq:posg}):  if the transmission
probabilities are exponentially small, then such terms would dominate the
denominator of~(\ref{eq:posg}), and the expression would no longer be valid.
To check that this does not happen, we write
   \eqn A = \left( \begin{array}{cccc}
     R(1+\alpha) & T & t & t \\
     T & R(1+\beta) & t & t \\
     t & t & R(1+\gamma) & T \\
     t & t & T & R(1+\delta)  \end{array} \right)
   \eqnend
where $\alpha$, $\beta$, $\gamma$ and $\delta$ are small compared to 1, but
not exponentially small.
For simplicity we ignore any asymmetries in the transmission coefficients
as they will be smaller than $T$ and $t$ and hence not qualitatitively
effect the conclusions of~(\ref{eq:posg}). We will also
ignore terms of order $t$
This leads to
   \eqn
       G = \frac{2e^2}{\hbar}
    \frac{(R^2-T^2)(1+  [(\alpha + \beta + \alpha\beta)/Y)]
         (1+  [(\gamma + \delta + \gamma\delta)/Y])}
         {2T(1  +
          [(1 + \alpha  + \beta + \alpha\beta +
                     \gamma + \delta + \gamma\delta)/2Y])}
   \eqnend
where $Y=1-T^2/R^2$
And we see that the asymmetries lead only to the exponentially small terms
in the denominator (shown in square brackets)
and hence do not effect our results.

In conclusion, we have studied the bend conductance of crossed
wire arrangements in the presence of Andreev scattering. By examing the
relative probabilities of transmission straight across the device and around
the corners we have been able to make qualitative statements about the
sign of the bend conductance, and have shown that if the superconductor
is confined to the area of the junction and the order parameter is
sufficiently strong that the low-energy quasi-particles are unable to
penetrate across the device then switching on the order parameter will
cause $G_{14,32}$ to reverse sign. For conventional superconductors,
typically $\Delta \sim 10^{-3}E_F$ at zero $T$ and $k_F^{-1} \sim$ a
lattice constant $\sim 10^{-10}\; m$. Hence the critical length scale above
which the conductance changes sign is of order $10^{-7}\; m$. This is
within the range of mesoscopic experiments, so that both of the
limits described in this letter should be experimentally observable.

The sign change predicted here should be compared to
that recently predicted for the longitudinal conductance
of a disordered wire~\cite{here:negative94}. In that case, the result depended
on
a relatively large Andreev scattering probability across the wire, and
hence the sign-change is only likely to be observable in the presence of
a magnetic field or magnetic impurities.  By contrast, for the
situation described by equation~(\ref{eq:posg}) the change of sign
arises because of the evanescent character of quasi-particle waves
below the superconducting energy gap, and hence should not require
magnetic effects to be observed. On the other hand, since scattering
around bends is highly sensitive to the presence of a magnetic
field,
 our results suggest that the conductance
superconducting crossed wires as a function of applied magnetic
field may have a rich structure.

This work has benefitted from financial support from the Ministry
of Defence, and from useful discussions with Colin Lambert and
Mark Jeffery.

\pagebreak

\pagebreak

{\centerline \bf List of Figures}

1. Crossed wire geometry. The shaded area is the region in which a
constant nonzero order parameter is assumed to exist when the device
becomes superconducting. The numbers label  the probes, $W$ is the
width of the junction and $L$ is the distance the superconducting
region extends beyond the junction.
Paths through the device $\gamma_1$, $\gamma_2$ and $\gamma_3$
 are marked, with respective lengths  inside the superconducting area
of $2L$, $4L$ and $(2L+W)$.


\begin{thebibliography}{10}
\bibitem{note}
{It should be noted that in order to
simplify our later analysis we have numbered the probes here in a different
order from that of ref. [5]. Also, eq. (1) contains an extra farctor of 2 to
alllow for spin degeneracy, not included in [5]. }
\end{thebibliography}
\end{document}